\documentclass[reprint,prl]{revtex4-1}

\pdfoutput=1

\usepackage[italian,english]{babel}
\usepackage{amssymb, amsmath}
\usepackage{graphicx}
\usepackage{mathrsfs,eufrak}

\makeatletter
\let\cat@comma@active\@empty
\makeatother

\begin{document}
%\title{Electric field effect on spin waves and on magnetization dynamics}
\title{Local spin Seebeck imaging with scanning thermal probe}

\author{Alessandro Sola}
\email{a.sola@inrim.it}
% \altaffiliation[Also at ]{Physics Department, XYZ University.}%Lines break automatically or can be forced with \\
\author{Vittorio Basso}
\author{Massimo Pasquale}

\affiliation{
	Istituto Nazionale di Ricerca Metrologica, Strada delle Cacce 91, 10135, Torino, Italy\\
}%

%\collaboration{MUSO Collaboration}%\noaffiliation

\author{Carsten Dubs}
% \homepage{http://www.Second.institution.edu/~Charlie.Author}
\affiliation{
	INNOVENT e.V., Technologieentwicklung, Pr\"ussingstrasse. 27B, 07745 Jena, Germany\\
}%

%\affiliation{
% Third institution, the second for Charlie Author
%}%
\author{Craig Barton}
\author{Olga Kazakowa}
\affiliation{
	National Physical Laboratory, Teddington TW11 0LW, United Kingdom\\
}%

%\collaboration{CLEO Collaboration}%\noaffiliation

\begin{abstract} In this work we present the results of an experiment to locally resolve the spin Seebeck effect in a high-quality Pt/YIG sample. We achieve this by employing a locally heated scanning thermal probe to generate a highly local non-equilibrium spin current.
	To support our experimental results, we also present a model based on the non-equilibrium thermodynamic approach which is in a good agreement with experimental findings. To further corroborate our results, we index the locally resolved spin Seebeck effect with that of the local magnetisation texture by MFM and correlate corresponding regions.
	We hypothesise that this technique allows imaging of magnetisation textures within the magnon diffusion length and hence characterisation of spin caloritronic materials at the nanoscale. \end{abstract}
 
%\begin{document}

\maketitle

\section{Introduction}
\label{sec:Introduction}

The visualisation of domain structure in magnetism and magnetic materials is paramount in aiding the understanding at the fundamental level and subsequent utilisation of such materials in real world applications. Hence, a significant effort is dedicated to the development of a variety of techniques which are suitable to study magnetic materials and magnetic domains.  Magnetic force microscopy (MFM) \cite{martin1987magnetic,saenz1987observation} is a scanning probe technique capable of sensing the force gradients induced by stray fields over the surface of a magnetic material at the nanoscale. Scanning electron microscopy with polarization analysis (SEMPA) \cite{scheinfein1990scanning} can also be used to visualize magnetic domains by monitoring the spin polarization of the secondary electrons interacting with the stray field of a sample under test. In order to visualize the local magnetization $M(r)$ (or the flux density $B(r)$) instead of the stray field $H(r)$, it is possible to use imaging techniques such as magneto-optic effects \cite{hubert2008magnetic} or Lorentz microscopy \cite{chapman1999transmission}. Electrons can also be used as a probe for the imaging of magnetic domains in electron holography \cite{tonomura1987applications}. The above techniques typically involve the characterization of magnetic properties confined to the surface or thin samples; the investigation of the domain structure in bulk materials, however, requires more complex experiments that involves neutron scattering \cite{halpern1941passage,pfeiffer2006neutron,grunzweig2008neutron,hilger2018tensorial} or X-ray spectroscopy \cite{streubel2015retrieving,donnelly2015element,dierolf2010ptychographic}.

The interaction between heat and non-equilibrium spin currents in magnetic materials represents an alternative approach to image magnetic domains. Analogous to standard thermoelectric effects, this interaction has been described as the thermal generation of driving power for electron spin, i.e. the spin Seebeck effect (SSE) \cite{uchida2008observation}. This effect involves pure spin currents which are produced by driving the system out of equilibrium through thermal gradients; by definition, they carry zero net charge and depend on the local magnetization of the material. The length scale $l_{\textrm{M}}$ that governs the SSE has been demonstrated to be of the order of micrometers \cite{kehlberger2015length,guo2016influence,basso2018non}.
\begin{figure*}[t]
	\centering
	\includegraphics[width=17.5cm]{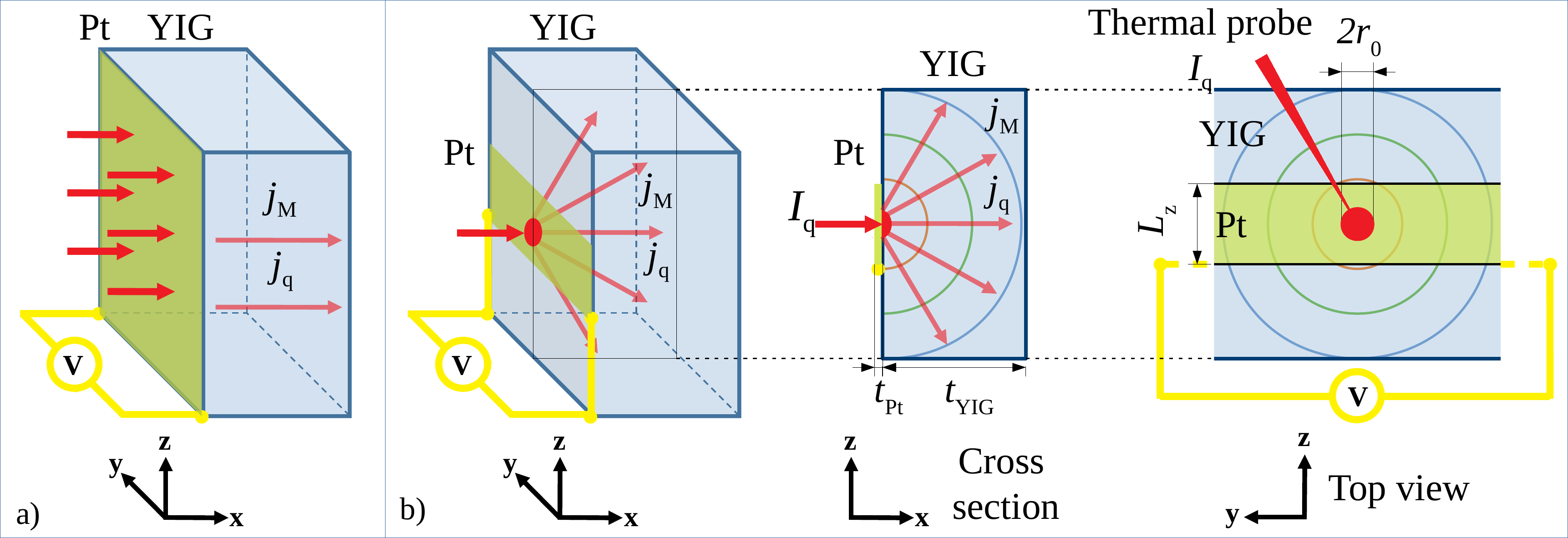}
	\caption{
		Longitudinal spin Seebeck measurement configurations. (a) Standard configuration with uniform heating (the red arrows at the Pt/YIG interface represent the hot side; the cold side is at the right-hand side of the YIG slab). (b) A local heating from the Pt side generates a magnetic moment current density $j_{\textrm{M}}$ in spherical symmetry. On the right side of the panel, the projections of the experimental scheme on the x-z and y-z planes are represented: the local heat current $i_{\textrm{q}}$ from the thermal probe spreads over a circle whose dimensions ($r_{0}$) limit the space resolution. For Pt/YIG sample that we investigated the dimensions are: $t_{\textrm{Pt}} = 5$ nm, $t_{\textrm{YIG}} = 0.5$ mm, $L_{\textrm{z}}^{Pt} = 150 \mu$m.
	} \label{fig1}
\end{figure*}
The SSE has been observed in magnetic insulator garnet ferrites \cite{uchida2010} such as the ferrimagnetic yttrium iron garnet Y$_{3}$Fe$_{5}$O$_{12}$ (YIG), or other ferrites \cite{uchida2010longitudinal,uchida2013longitudinal,meier2013thermally}. A typical spin Seebeck device can be formed by creating a bilayer of a magnetic material and a thin metallic film with high spin-orbit coupling like platinum or tungsten both of which are paramagnetic heavy metals. This second layer acts as the spin detector thanks to the inverse spin Hall effect (ISHE) that enables a spin-charge conversion at the interface \cite{Saitoh2006}. The most useful configuration is the longitudinal spin Seebeck effect (LSSE) \cite{uchida2010observation,uchida2014longitudinal,meier2015longitudinal} that corresponds to the generation of a spin current parallel to the temperature gradient both of which are orthogonal to the local magnetisation direction which is in the sample plane. The majority of experiments typically refer to spin Seebeck samples that are uniformly heated over the whole sample and are in a uniform magnetic state, i.e. at or near saturation. In these conditions, it is possible to compare the spin Seebeck characteristics to the magnetization loop of the sample; however it is not possible to resolve the contributions to the spin Seebeck signal coming from regions where the orientation of the magnetization differs from the average, i.e. in a multidomain state. To resolve magnetic domains, it is necessary to go beyond the experimental configuration previously described in favour of a locally injected heat current, as described in Figure \ref{fig1}. The dependence of a local spin Seebeck signal on the local magnetization has been observed by scanning a laser beam on a Pt/YIG structure \cite{weiler2012local} and in a time-resolved configuration \cite{roschewsky2014time,bartell2017imaging,jamison2019long}. By taking account of the experimental geometry the same set-up can be adopted for the measurement of the time resolved anomalous Nernst effect \cite{weiler2012local,bartell2015towards}.
In this work we present for the first-time local measurements of the SSE using a thermal AFM probe as local source of heat and non-equilibrium spin currents. We employed a high-quality bulk YIG single crystal with a Pt strip lithographically defined onto the surface as our spin detector. We demonstrate that the measured effect is unambiguously the local spin Seebeck through a series of tests; by varying the heating power and the vector of the externally applied magnetic field. We interpret and support our observations with a thermodynamic description of the generation of the local magnetic moment current. This model describes quantitatively the geometry of local heat current as a circular heat source below the thermal probe whose size is larger than the cross-section of the tip due to the non-zero thermal conductivity at the Pt/YIG interface. The diameter of the heat source that we observed by employing our model is $\sim 2.8$ $\mu$m.

\section{Spin Seebeck effect by uniform and local heating}
\label{sec:Local}

\begin{figure*}[t]
	\centering
	\includegraphics[width=18cm]{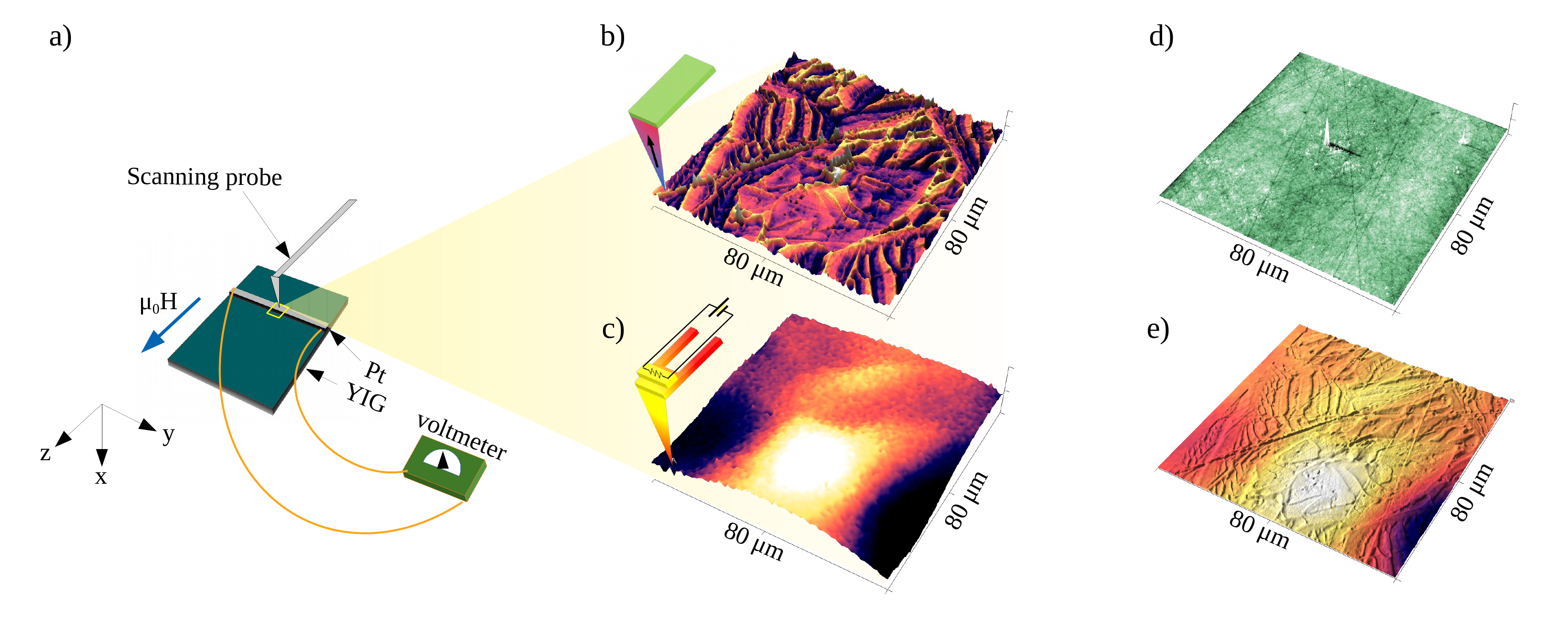}
	\caption{
		Experimental set-up for the local SSE measurements. (a) Macroscopic schematic representation of the Pt/YIG sample with the nanovoltmeter connected at the edges of the Pt strip. The subsequent images (b-e) are examples of measurements obtained from a 80 $\mu$m $\times$ 80 $\mu$m area of the Pt film with an applied magnetic field of $\sim 8$ mT. (b) MFM image of the same area, (c) spin Seebeck voltage originating from the scanning thermal probe, (d) AFM image that is obtained both from MFM and local spin Seebeck map, (e) 3D image obtained by overlapping of MFM and spin Seebeck map, taking into account the information on their mutual shift provided by the AFM images.
	} \label{fig2}
\end{figure*}

Figure \ref{fig1} shows two variations of a spin Seebeck experiment, both demonstrating that the measured signal due to the inverse spin Hall effect (represented by the yellow wires in Figure \ref{fig1}) corresponds to the open circuit voltage across the Pt film. The first one (Figure \ref{fig1} (a)), represents the “classical” experimental geometry reported by the majority of experimental results. This provides a uniform distribution of the thermally-generated spin current as consequence of the heat current across the whole surface of the sample. The proportionality between the voltage gradient $\nabla_{y} V_{\textrm{SSE}}$ generated along the Pt film due to the SSE and the heat current $I_{\textrm{q}}$ passing through the sample along the cross-section $A$ is described by the following expression:
\begin{equation}
	\frac{\nabla_{y} V_{\textrm{SSE}}}{{I_{\textrm{q}}}}\kappa A = \theta_{\textrm{SH}} \left(\frac{\mu_{\textrm{B}}}{e} \right) \frac{v_{\textrm{YIG}} l_{\textrm{YIG}} \epsilon_{\textrm{YIG}}}{v_{\textrm{p}}} \frac{1}{t_{\textrm{Pt}}}
	\label{DeltaV_uniform}
\end{equation}

\noindent where $\mu_{\textrm{B}}/e$ is the ratio between the Bohr magneton and the elementary charge, $\kappa$ is the thermal conductivity of YIG.
The other parameters inside Eq. \ref{DeltaV_uniform} are the spin Hall angle $\theta_{\textrm{SH}}$, the absolute thermomagnetic power coefficient $\epsilon_{\textrm{YIG}}$, the magnon diffusion length $l_{\textrm{YIG}}$ and the thickness of the Pt film $t_{\textrm{Pt}}$.
The intrinsic magnetic moment conductance of YIG is represented by $v_{\textrm{YIG}} = l_{\textrm{YIG}}/\tau_{\textrm{YIG}}$ where $\tau_{\textrm{YIG}}$ is the magnon mean scattering time.
The parameter $v_{\textrm{p}}$ represents the magnetic moment conductance per unit surface area of the Pt/YIG bilayer. This quantity depends on the intrinsic conductances of YIG and Pt, on the ratio between the thickness and the magnon diffusion length, for each layer. We derived the expression of $v_{\textrm{p}}$ from a thermodynamic description of the magnetic moment currents generation \cite{basso2018spin}. 
It is important to note that the geometry shown in Figure \ref{fig1} and modelled by Eq. \ref{DeltaV_uniform} does not allow to resolve the spatial distribution of the underlying magnetic structure, since the spin Seebeck voltage results from an averaged contribution of regions with different magnetization.
Because of this, the experiments performed in this geometry are usually conducted at magnetic saturation or follow the magnetization hysteresis loop.

However, in the second experimental setup (Figure \ref{fig1} (b)), which shows the effect of the heated AFM probe in contact with the Pt surface, we can selectively generate the heat current injected through a point of the Pt surface and propagating with spherical symmetry in the volume. In this arrangement, the thermally generated spin current is assigned to a locally limited SSE, which is generated at the point of thermal contact, represented by the red dot in Figure \ref{fig1} (b). As in the standard configuration the spin Seebeck voltage depends on the average magnetization, in the configuration with local heating the effect scales with the magnetization of a region whose size is determined by the locally heated volume, allowing resolved measurements of the domain distribution within the sample. To describe this experimental configuration, we took into account the local magnetization $\mathbf{m}$ as a unit vector. This leads to an expression of the spin Seebeck voltage, which originates from a circle with radius $r_{0}$ on the top of the sample, into which a heat current $I_{\textrm{q}}$ is injected. In this way it is possible to rewrite Eq. \ref{DeltaV_uniform} as follows:
\begin{equation}
	\frac{\nabla_{y} V_{\textrm{SSE}}}{{I_\textrm{q}}}2\pi r_{0}^{2}k = \theta_{\textrm{SH}} \left(\frac{\mu_{\textrm{B}}}{e} \right) \frac{v_{\textrm{YIG}} l_{\textrm{YIG}} \epsilon_{\textrm{YIG}}}{v_{\textrm{p}}} \frac{1}{t_{\textrm{Pt}}}
	\label{DeltaV_result}
\end{equation}

\noindent where the dependence on the radius of the heated region $r_{0}$ (red circle in Figure \ref{fig1} (b)) is highlighted.
However, the value of $V_{\textrm{SSE}}$ in Eq. \ref{DeltaV_result}, and in particular the voltage difference $\Delta V_{\textrm{SSE}}$ is the one that would be measured at the two ends of the heated region, limited by $r_{0}$. Since this quantity is not accessible by the experiment, it is necessary to rescale the value of $\Delta V_{\textrm{SSE}}$, taking into account the lateral dimensions of the whole Pt film that works as the spin-voltage detector. Here we use an approximation criterion and consider the Pt film as an electric circuit formed by a voltage source ($V_{\textrm{SSE}}$), two resistances in series whose sum is proportional to $(L_{\textrm{y}}-2r_{0})/2r_{0}$ and one resistance in parallel whose value is proportional to $L_{\textrm{y}}/(L_{\textrm{z}}-2r_{0})$. Assuming that $2r_{0}\ll L_{\textrm{y}}$ and $2r_{0}\ll L_{\textrm{z}}$, we can represent the experimental values of the spin Seebeck voltage as a rescaled value of $\Delta V_{\textrm{SSE}}$
\begin{equation}
	\Delta V_{\textrm{SSE,exp}} \simeq \Delta V_{\textrm{SSE}} \frac{2r_{0}}{L_{\textrm{z}}}
	\label{DeltaV_rescaled}
\end{equation}

\noindent The expression of $\Delta V_{\textrm{SSE,exp}}$ can be used to interpret the experimental data of the local SSE.
By using the parameters included in Eq. \ref{DeltaV_result} to determine the value of $V_{\textrm{SSE}}$ at the right of Eq. \ref{DeltaV_rescaled}, it is possible to obtain a value of $r_{0}$ that can be considered as the space resolution of the spin Seebeck imaging.

\section{Experimental methods}
\label{sec:Experimental}

The measurements of the local SSE were performed on a high quality YIG single crystal plate ($L_{y} = 4.95$ mm, $L_{z} = 3.91$ mm and thickness $t_{YIG} = 0.545$  mm). Both crystal surfaces were polished to be optically flat  ($R_{q} = 0.4$ nm obtained by AFM). A 150 $\mu$m wide Pt strip was sputtered onto one side of the YIG crystal along the y direction, where the z direction defines the stripe width. The thickness of the Pt film $t_{YIG}$ was chosen equal to be approximately $5$ nm in accordance with \cite{castel2012platinum,wang2014scaling}. The measurement technique for the uniform heat current, i. e. the classical configuration, has been described elsewhere \cite{sola2019experimental} for the same sample. The local heat current was generated using a nano thermal analysis probe (NanoTa probe); this consists of a micropatterned AFM cantilever probe that allows a current to be driven around the cantilever resulting in Joule heating that propagates to the tip apex. We approximated the temperature of the probe using a series of polymer test samples with known glass transition temperatures. We also expected an offset in our estimated temperatures due to the respectively higher thermal conductivity of the Pt used in our studies as compared to the test polymers. With this procedure it was possible to obtain an approximate relationship between the probe power voltage and the temperature difference applied to the sample in this geometry. Moreover, we observed a stable spin Seebeck signal when reverting the heating voltage thus verifying absence of electric interference from the nanoTA cantilever. This setup allowed us to compare a local domain map obtained by MFM with a local SSE map and correlate the two data sets. The enlarged area in Figure \ref{fig2} (a) represents one of the specific regions investigated in the experiments. From the data in Figure \ref{fig2} (b,c) it is possible to spatiality map the SSE voltage at the location of the thermal probe in contact with the Pt film. 
This makes it possible to correlate SSE and MFM and draw qualitative conclusions. Figures \ref{fig2} (b) and (c) show a stray field gradient at the YIG surface obtained by MFM (Figure \ref{fig2} (b)) and a SSE map obtained by nanoTA (Figure \ref{fig2} (c)). The experiment was structured as follows: first we performed a set of nanoTA measurements with a saturating magnetic field in both directions ($\pm M_{\textrm{s}}$), achieved using a small neodymium magnet adjacent to the sample. This corresponds to a standard SSE experiment. The second set of measurements was performed at a magnetic field where a domain structure could be observed using MFM. Keeping the applied field constant, the nanoTA measurements were performed at different heating power levels of the thermal probe and at several points of the sample surface including the transition from the Pt strip to the bare YIG surface.

\section{Results and discussion}
\label{sec:Results}
\subsection{\label{sec:with_applied}Local spin Seebeck effect at magnetic saturation}

\begin{figure}[t]
	\centering
	\includegraphics[width=8cm]{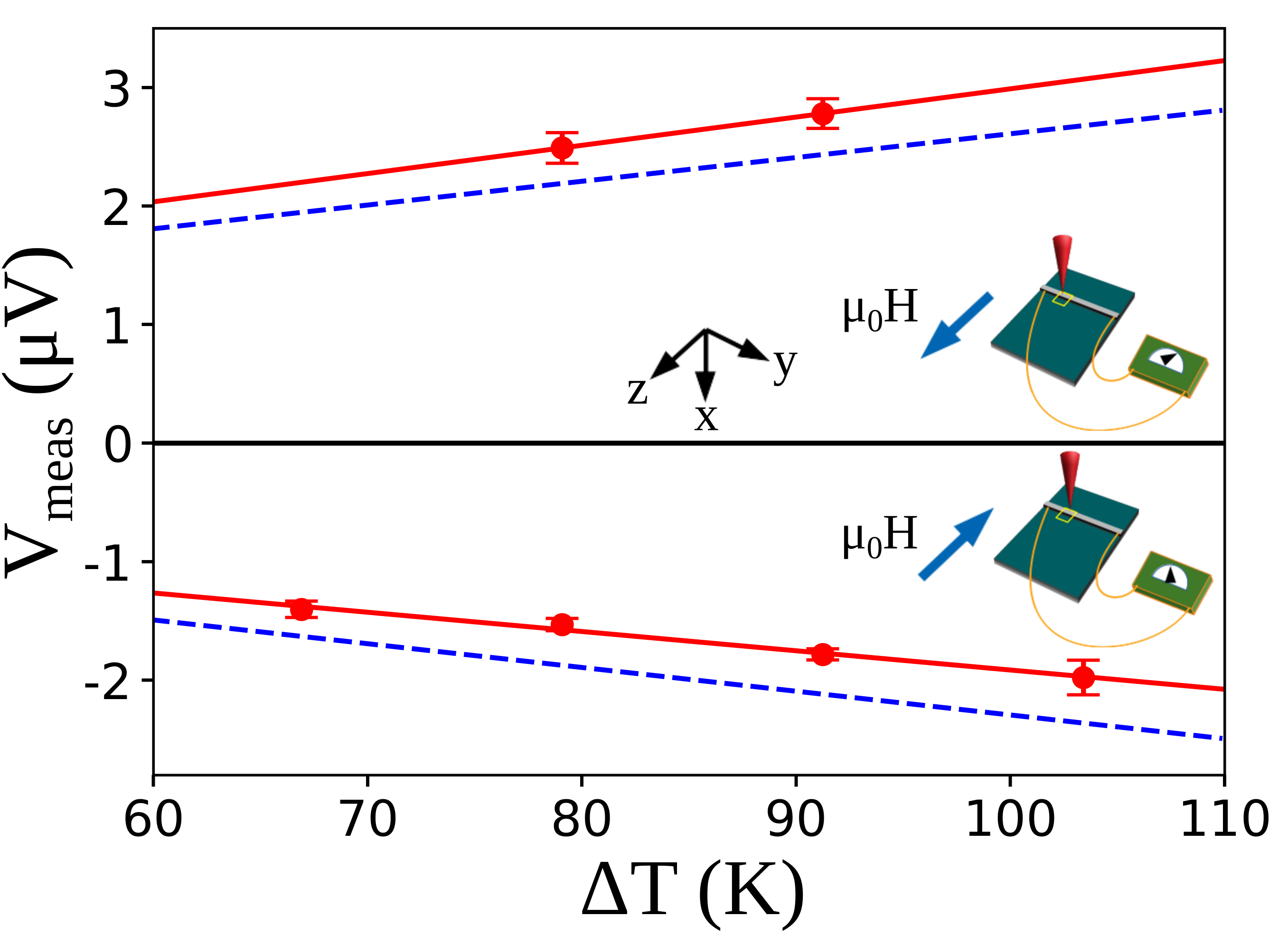}
	\caption{
		Measured voltage as consequence of the local heating of the sample at magnetic saturation: spin and ordinary Seebeck voltage as function of the temperature differences (red point). Fitting lines after the compensation of the ordinary Seebeck component (blue dashed lines). The direction of the saturating magnetic field is shown as a sketch in each panel. 
	} \label{fig3}
\end{figure}

We first investigated the local SSE of the Pt/YIG bilayer structure at magnetic saturation. This experiment was performed with the field aligned within the plane of the sample and perpendicular to the long axis of the Pt strip (Figure \ref{fig2} (a)). The experimental data points in Figure \ref{fig3} show the SSE data for each realized temperature differences $\Delta$T.
\noindent Here, we averaged the signal that originated from the local heating of a 40 $\mu$m $\times$ 20 $\mu$m area of the Pt surface and the experimental uncertainty was evaluated from the standard deviation of these data sets. The dependency between the total measured voltage, including the spin Seebeck contribution and $\Delta$T can be described by the following relation:
\begin{equation}
	V_{\text{meas}}=\pm C_{\text{SSE}}\Delta T + C_{\text{OSE}}\Delta T + V_0
	\label{Fitting_eq}
\end{equation}

\noindent where $V_{\text{meas}}$ corresponds to the average voltage recorded as a consequence of scanning a given area and $\Delta$T is the difference between the probe temperature and the room temperature. The voltage $V_{\text{meas}}$ contains the following contributions: the first one ($C_{\text{SSE}}\Delta T$) corresponds to the spin Seebeck voltage $V_{\textrm{SSE,exp}}$ of Eq. \ref{DeltaV_rescaled}; the sign of the spin Seebeck coefficient $\pm C_{\text{SSE}}$ depends on the direction of the magnetisation.
The second contribution is the ordinary Seebeck effect ($C_{\text{OSE}}$); for our work, this is a spurious component that does not depend on the magnetic configuration of the sample. The $C_{\text{OSE}}$ component derives from the contact between the different metals used to electrically connect the sample (silver paint for bonding the platinum strip at the edges of the thin film). This is due to a small transverse heat loss through the electric contacts associated to small geometric asymmetry. Such an artefact can affect the interpretation of spin-caloritronic measurements and has been previously described for both the spin Seebeck and spin Peltier effect \cite{sola2019experimental} data. The last contribution of Eq. \ref{Fitting_eq} is an offset voltage $V_0$ that originated from the circuit resistance.
The voltage due to the Seebeck effects, both spin and ordinary effects, can be plotted as function of $\Delta$T, according to Eq. \ref{Fitting_eq}. From the difference between the absolute values of the slopes obtained by the linear fits shown in Figure \ref{fig3} (red lines), a clear distinction can be made between the ordinary Seebeck and the spin Seebeck components. After the compensation of the ordinary Seebeck component, we present the spin Seebeck data by blue dashed lines in Figure \ref{fig3}, whose coefficient is $C_{\text{SSE}} = 2\cdot 10^{-8}$ VK$^{-1}$.

\subsection{\label{sec:with_applied}Local spin Seebeck effect at intermediate magnetization obtained with $\sim 8$ mT}

\begin{figure}[h]
	\centering
	\includegraphics[width=8cm]{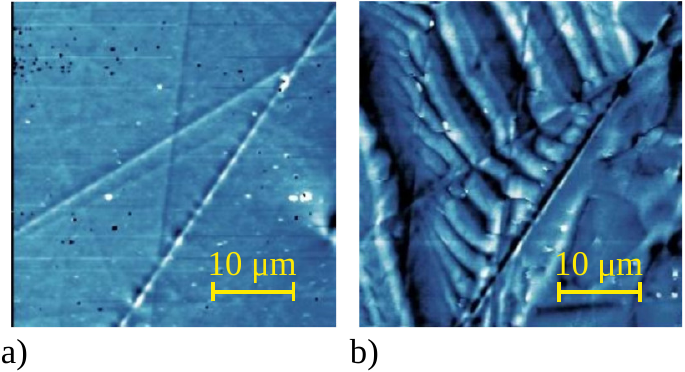}
	\caption{
		MFM micrograph of the same area of the Pt/YIG surface at magnetic saturation (left panel) and at $\sim 8$ mT (right panel).
	} \label{fig4}
\end{figure}

In the second step, we repeated the measurements with a lower applied magnetic field in order to induce a reduction in the magnetostatic energy of the sample and introduce a domain structure where we could distinguish several magnetization areas from the MFM images. We applied the magnetic field at lowered level by distancing the neodymium magnet from the sample and we measured its value by positioning an Hall probe in place of the sample. The measured value for the applied magnetic field was approximately 8 mT.
An example of two MFM images of the sample at magnetic saturation and with an applied field of $\sim 8$ mT obtained from the same area is represented in Figure \ref{fig4}. 

\begin{figure}[t]
	\centering
	\includegraphics[width=6cm]{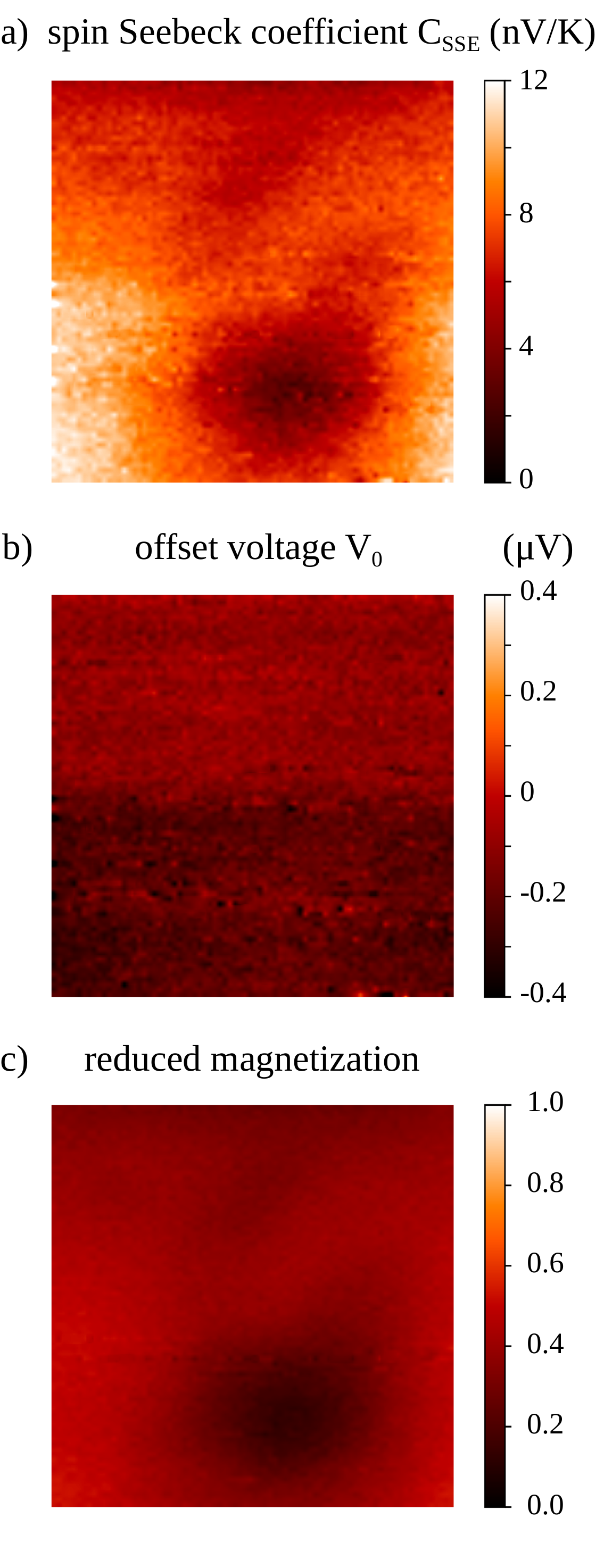}
	\caption{
		spin Seebeck maps of one 80 $\mu$m $\times$ 80 $\mu$m area of the Pt/YIG sample surface obtained with a local heating generated by 0.5, 1, 1.5, 1.8 and 2.2 V on the heater. (a) Spin Seebeck coefficients measured at each pixel of the map. (b) Map of the spurious offset voltage component and (c) map of the ratios between the local spin Seebeck voltage and the corresponding quantity at magnetic saturation; the colour scale in (c) represents the magnetization as percentage of the magnetic saturation.} \label{fig5}
\end{figure}

\begin{figure*}[t]
	\centering
	\includegraphics[width=18cm]{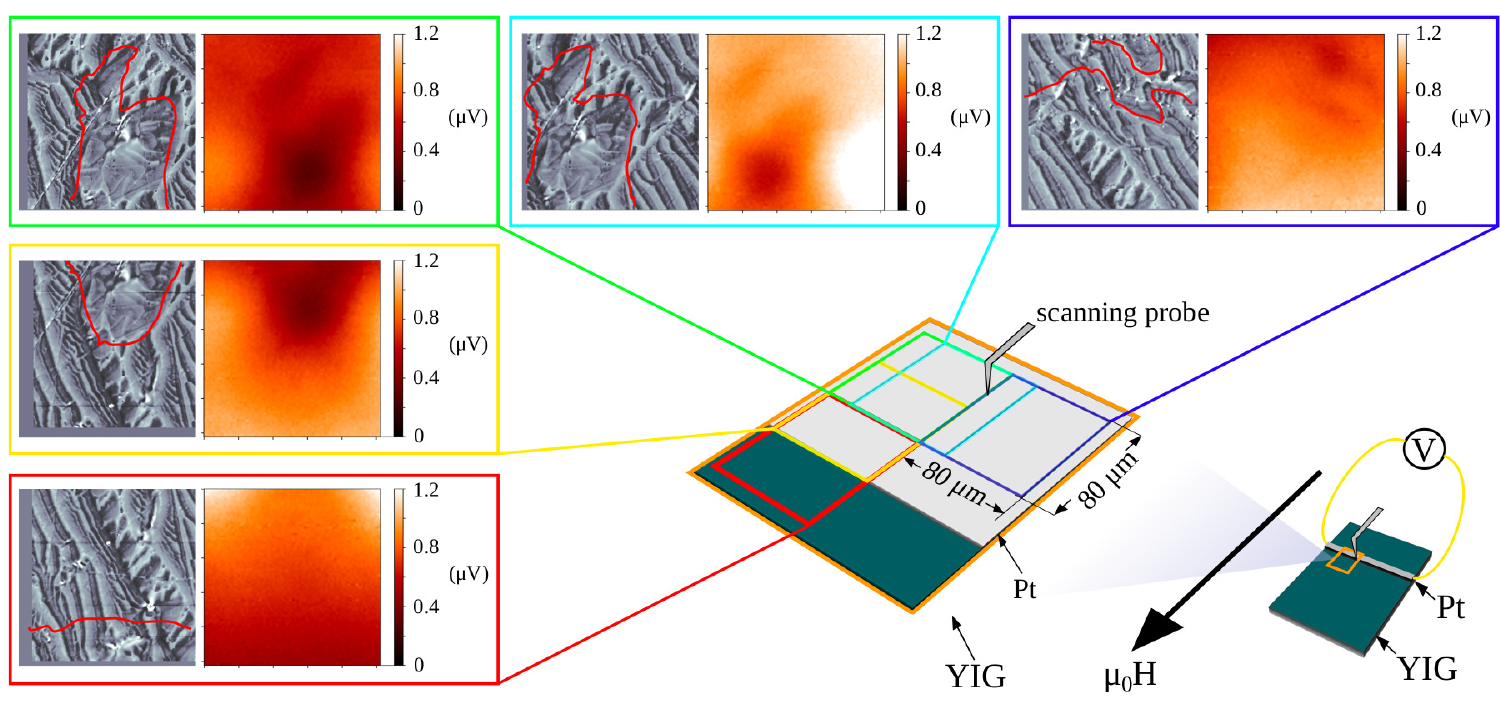}
	\caption{
		Local spin Seebeck voltage maps (right columns) obtained at $\sim 8$ mT with 1.5 V at the heater, shown together with the MFM micrographs (left columns). The five pairs of maps correspond to the areas of the sample represented in the scheme, where the edge of the Pt film is shown, together with the direction of the applied magnetic field.
	} \label{fig6}
\end{figure*}

\noindent Having applied $\sim 8$ mT, we scanned the thermal probe over a 80 $\mu$m $\times$ 80 $\mu$m in the same locations as that of the MFM, at different values of $\Delta$T in order to analyse the voltage that arises form the local SSE.
We used five values of heating power on the thermal probe that led to the corresponding temperature differences $\Delta$T at each area investigated. With these data sets we were able to determine the SSE dependence on heating power and extract the SSE voltage for each pixel of the data set. First we removed the ordinary Seebeck component, as described for the sample at magnetic saturation; this procedure gives the spin Seebeck coefficient $C_{\text{SSE}}$ and the measurement offset $V_0$, represented by the slopes and the intercepts of the blue lines in Figure \ref{fig3}.
Here we extract the values of these two parameters for the sample at intermediate magnetization by performing a linear fit of the voltage values ($V_{\text{meas}}$) represented by each pixel as function of the temperature difference. We can now build two maps that represent the parameters of these linear fits: the map of the spin Seebeck coefficients $C_{\text{SSE}}$ (Figure \ref{fig5} (a)) and the map of intercepts $V_0$ (Figure \ref{fig5} (b)).
Since we have already measured the upper limit of the spin Seebeck signal i.e. the value at magnetic saturation, we can use this value to normalize our results with $\sim 8$ mT which is represented as a percentage of the maximum signal at saturation. Figure \ref{fig5} (c) shows an example of this procedure recorded at the maximum temperature difference; the colorbar quantifies the level of magnetization between zero and the values at saturation, represented as the range $[0 - 1]$. 
From the data in Figure \ref{fig5} (a), we found that the ratio between the spin Seebeck values and the temperature difference (i.e. the spin Seebeck coefficient) has some non-zero positive values in some regions and decreases to zero in the round central region of the map. By adapting the definition of the spin Seebeck coefficient that is formulated for the saturated sample to the case of intermediate magnetization, we can deduce what follows. In some areas the magnetization was oriented to provide a larger spin Seebeck signal with respect to the areas where we observed a lower signal, reasonably due to a change in the orientation of the magnetization.
On the contrary, the map of the intercepts in Figure \ref{fig5} (b) is considerably flatter, indicating that the electric offset was rather steady over the investigated area. 

\subsection{\label{sec:comparison}Comparison between MFM and local spin Seebeck images}

The third step of this experiment involved scanning different regions of the sample, at a fixed heating voltage, with the same applied magnetic field as previously used ($\sim 8$ mT). The motivation is a qualitative comparison of the spin Seebeck maps with MFM micrographs. We tested the hypothesis that the contrast of the local spin Seebeck map is related to the local magnetization through the local spin Seebeck signal. Scanning of large areas allowed for overlapping neighbouring data sets and while recorded them accordingly; the corresponding MFM micrographs and the local spin Seebeck voltage maps are reported in Figure \ref{fig6}.
The signal of the local spin Seebeck maps was processed using the Gwyddion data analysis software \cite{nevcas2012gwyddion}. First we focused on the presence of some sharp voltage spikes; these have been manually corrected with an interpolation of the error-free pixels surrounding the spike. The second data process concerned the artefacts that usually appears because of the line by line acquisition. We corrected the voltage shift that rise between neighbouring horizontal lines by minimizing the median of height differences, between vertical neighbouring pixels \cite{klapetek2018quantitative}.
From Figure \ref{fig6} we observe that the MFM micrographs can provide a great variety of details due to the high resolution, whereas the local SSE maps have a lower spatial resolution. Furthermore, we have to keep in mind that the underlying physical phenomena are different for both experiments: the MFM signal depends on the stray fields emanating from the sample surface whereas the local SSE depends on the magnetization of the sample and the integrated effect over hot spot provided by the probe. Nevertheless, there is a clear correspondence between regions and features in the MFM micrographs and the regions of local minima of the local spin Seebeck maps. To make the correlation more visible, we traced the contour lines extracted by the spin Seebeck maps above the MFM micrographs in Figure \ref{fig6}. For the three images on the left, we selected the contour lines corresponding to $0.6$ $\mu$V on the spin Seebeck map, while for the two images on the right we selected different values since we did observe a drift on the offset voltage. For this reason, the two areas labelled by the green and light-blue frames appear on a slightly different voltage scale, but nevertheless it is possible to distinguish the shape of the local minimum in the expected position, according to the $20$ $\mu$m shift between the two areas.
The two micrographs that refer to the area labelled by the red frame in Figure \ref{fig6} represent a measurement across the edge of the Pt strip, cutting the frame horizontally. In the lower half of the SSE micrograph (the region not covered by the Pt), the spin Seebeck voltage decreases in agreement with the local generation of a spin current from the bare YIG that is not detected by the Pt film. Finally, it is possible to hypothesise the level of the local magnetization of the sample. This was achieved by averaging the values reported in the local spin Seebeck maps of Figure \ref{fig6} and then normalising these values with the spin Seebeck coefficient according to the procedure adopted for Figure \ref{fig5} (c). By considering this approach, we obtained a value of the average magnetization corresponding to $0.4$ times the magnetic saturation. This value is in agreement with the measurement performed with the Hall probe (8 mT), knowing that over 20 mT the sample saturates (see supplementary informations of ref. \cite{sola2019experimental}).
This value tells us that the sample was far from the magnetic saturation but above the level of magnetization at which the magnetic-field dependence of the SSE deviates from the bulk magnetization curve \cite{uchida2015intrinsic,kalappattil2017roles}, as reported by other experiments performed on bulk samples \cite{uchida2010observation,kikkawa2013longitudinal,uchida2013longitudinal,kikkawa2013separation,uchida2010longitudinal,kikkawa2015critical}.

\subsection{\label{sec:reslution}Spatial resolution of the spin Seebeck image}

Finally, we comment on the resolution of the local SSE measurement technique. We consider the radius of the hot spot ($r_{0}$ in in Eq. \ref{DeltaV_result}) as a more realistic limit to the resolution, compared to the cross-section of the probe. Starting from the thermodynamic description of the SSE, we have derived the expression of the spin Seebeck voltage difference as function of the heat current injected by the heated AFM probe. We also highlighted the need to rescale this expression according to the geometry of the experiment, as represented by Eq. \ref{DeltaV_rescaled}. By replacing the expression of the spin Seebeck voltage $\Delta V_{\text{SSE}}$ (Eq. \ref{DeltaV_result}) inside the expression of the rescaled signal we could derive the following representation of the experimental results: 
\begin{equation}
	\frac{\Delta V_{\textrm{SSE,exp}}}{\Delta T} = \frac{4 r_{0}}{L_{\textrm{z}}}
	\theta_{\textrm{SH}} \left(\frac{\mu_{\textrm{B}}}{e} \right) \frac{v_{\textrm{YIG}} l_{\textrm{YIG}} \epsilon_{\textrm{YIG}}}{v_{\textrm{p}}}
	\frac{1}{t_{\textrm{Pt}}}
	\label{DeltaV_resolution}
\end{equation}

\noindent where the parameters that represent the properties of YIG have been chosen according to the semi-infinite approximation where the thickness of the YIG is larger than that of the magnon diffusion length and $L_{\textrm{z}} = 150$ $\mu$m is the width of the Pt strip.
In Eq. \ref{DeltaV_resolution} the heat current $I_{\text{q}}$ that appears in Eq. \ref{DeltaV_result} has been written as the temperature difference $\Delta T$, by knowing the thermal conductivity $\kappa$ and the geometrical constrains on the YIG layer.
By using the experimental data from a previous spin Seebeck study on the same sample \cite{sola2019experimental}, we can use the following experimental value for the YIG.
\begin{equation}
	\frac{v_{\textrm{YIG}} l_{\textrm{YIG}} \epsilon_{\textrm{YIG}}}{v_{\textrm{p}}} = - 4.6 \times 10^{-10}
	\mbox{(V/K)(m/s)$^{-1}$}
	\label{YIG_parameters}
\end{equation}

\noindent The spin Hall angle $\theta_{\textrm{SH}}=-0.1$ refers to the current of magnetic moments which has the opposite sign with respect to a definition of the spin Hall angle based on the spin current \cite{sola2019experimental}.
By replacing the value of the spin Seebeck coefficient obtained ($C_{\text{SSE}} = 2\cdot 10^{-8}$ VK$^{-1}$) in place of $\Delta V_{\textrm{SSE,exp}}/\Delta T$ in Eq. \ref{DeltaV_resolution}, we obtained a value of $r_{0} = 1.4$ $\mu$m. This value of $r_{0}$ is in reasonable agreement with the lateral point-spread function of a heating laser presented by Bartell et. al. \cite{bartell2017imaging} with a FWHM of the hot spot of $0.606$ $\mu$m, obtained with an optical laser power of $0.6$ mW.

\section{Conclusions}
\label{sec:Conclusions}

In summary, we have employed a scanning probe microscopy technique to locally inject heat currents in a Pt/YIG bilayer structure. We observed a spatially-resolved voltage response dependent on the location of the heated probe and the local magnetisation state which we unambiguously attribute to the SSE. 
This allows for the first time to obtain locally resolved spin Seebeck measurements, which we map spatially and compare qualitatively with MFM micrographs obtained on the same scanned regions. We have discussed the measured signals using a thermodynamic description in spherical coordinates. Furthermore, we have derived the spatial resolution of the local spin Seebeck measurements which is of the order of few micrometres. Local spin Seebeck imaging represents an innovative tool for the investigation of novel spin caloritronic materials. In particular, it provides a significant step forward for the analysis of bulk magnetic structures, compared to surface characterization techniques, as the signal originates from the bulk at a distance that can be considered equivalent to the magnon diffusion length. Additionally, this experimental technique allows us to image the magnetisation structure of samples where tip-sample interaction could result in irreversible changes of the sample state during the imaging process, thus providing a non-perturbative imaging tool. Moreover, this technique could pave the way to new concepts of scanning probe microscopy, inspired by spintronics and spin-caloritronics. For example, the development of V-shaped Pt probes as a point-contact ISHE detector, or the scanning thermal microscopy (SThM) as a probe for spatial magnetic imaging using the spin Peltier effect.

\begin{acknowledgments}
	The authors thank the EMRP Joint Research Projects 15SIB06 NanoMag for financial support. In particular, A. S. thanks the Researcher Mobility Grant 15SIB06-RMG3. The EMRP is jointly funded by the EMRP participating countries within EURAMET and the European Union.
	
	We thank Dr. Vladimir Antonov, Royal Holloway, University of London (RHUL) for Pt deposition.
	
	A. S. thanks Marco Coisson for useful discussions and preliminary MFM measurements.
	
	C. D. thanks R. Meyer and B. Wenzel for technical assistance.
\end{acknowledgments}

%\bibliography{00BibQ}
%\bibliographystyle{plain}
%\bibliographystyle{unsrt}

\end{document}